\documentclass{SCIS2018}

\newcommand \Gaussian{\textit{Gaussian} }
\newcommand \gray{\textit{Gray} }
\newcommand \Gray{\textit{Gray} }
\newcolumntype{V}{!{\vrule width 1pt}}

\begin{document}

\ArticleType{RESEARCH PAPER}
\Year{2018}
\Month{January}
\Vol{61}
\No{1}
\DOI{}
\ArtNo{}
\ReceiveDate{}
\ReviseDate{}
\AcceptDate{}
\OnlineDate{}

\title{Polar-Coded Forward Error Correction for MLC NAND Flash Memory}{Polar FEC for NAND flash memory}

\author[1,2,3]{Haochuan Song}{}
\author[4]{Frankie Fu}{}
\author[4]{Cloud Zeng}{}
\author[5]{Jin Sha}{}
\author[2,3]{Zaichen Zhang}{}
\author[3]{\\Xiaohu You}{}
\author[1,2,3]{Chuan Zhang}{chzhang@seu.edu.cn}

\AuthorMark{Song H}

\AuthorCitation{Song H, Fu F, Zeng C, Sha J, Zhang Z, You X, Zhang C}


\address[1]{Lab of Efficient Architectures for Digital-communication and Signal-processing (LEADS)}
\address[2]{Quantum Information Center of Southeast University}
\address[3]{National Mobile Communications Research Laboratory, Southeast University, Nanjing {\rm 211189}, China}
\address[4]{Lite-On Technology Corporation, Guangzhou {\rm 510000}, China}
\address[5]{School of Electronic Science and Engineering, Nanjing University, Nanjing {\rm 210046}, China}

\abstract{With the ever-growing storage density, high-speed, and low-cost data access, flash memory has inevitably become popular. Multi-level cell (MLC) NAND flash memory, which can well balance the data density and memory stability, has occupied the largest market share of flash memory. With the aggressive memory scaling, however, the reliability decays sharply owing to multiple interferences. Therefore, the control system should be embedded with a suitable error correction code (ECC) to guarantee the data integrity and accuracy.
We proposed the pre-check scheme which is a multi-strategy polar code scheme to strike a balance between reasonable frame error rate (FER) and decoding latency. Three decoders namely binary-input, quantized-soft, and pure-soft decoders are embedded in this scheme.
Since the calculation of soft log-likelihood ratio (LLR) inputs needs multiple sensing operations and optional quantization boundaries, a $2$-bit quantized hard-decision decoder is proposed to outperform the hard-decoded LDPC bit-flipping decoder with fewer sensing operations. We notice that polar codes have much lower computational complexity compared to LDPC codes. The stepwise maximum mutual information (SMMI) scheme is also proposed to obtain overlapped boundaries without exhausting search. The mapping scheme using \Gray code is employed and proved to achieve better raw error performance compared to other alternatives. Hardware architectures are also given in this paper.}


\maketitle

\section{Introduction}\label{sec:intro}
Nowadays, the ever-developing digital technologies enable us to achieve extremely high communication speed. However, traditional hard disk drive (HDD) can no longer meet the throughput and latency requirements of most state-of-the-art application scenarios. To this end, NAND flash memory, which is of lower access time, higher compactness, and less noise has become increasingly popular for storage market \cite{li2010improving,Kim2012low}.

The past decade has witnessed the steady price fall of flash memory and is expecting further price-drop in the future \cite{Grupp2012bleak,Tutorial2013}. This trend has enabled solid state drive (SSD), which is mainly based on NAND flash memory, to occupy a large share of both business and consumer markets.

\subsection{Challenges and motivation}
As the required storage density increases, most NAND flashes consider to store $4$ bits in a single cell \cite{multi1,multi2,multi3,multi4}, which results in worse raw error performance. Therefore, powerful forward-error correction (FEC) methods are required, and voluminous researches on conventional error correction code (ECC) schemes for NAND flash memory emerge \cite{Xijia2015,kim2012performance,dong2011use,chen2008error,cui2014multilevel}. Recently, low-density parity-check (LDPC) codes have been considered. To balance performance and complexity, hybrid scheme combining hard decoder and soft decoder is always employed. However, the accepted soft decoders such as min-sum and belief-propagation (BP) suffer from high complexity. Identifying an alternative code might serve as a solution.

Recently, polar codes \cite{Arikan09} have shown capacity-achieving performance and reasonable complexity \cite{zhang2012reduced,zhang2013low}. Besides its good performance over binary-input discrete memoryless channels (B-DMCs), $N$-bit polar code's encoding and decoding complexity is as low as $O(N\log N)$, which is much lower than that of LDPC code. Consequently, polar codes have been selected as the control channel code for the enhanced mobile broadband (eMBB) scenario by 3GPP \cite{3gpp}. Inspired by few existing literature \cite{Li2015polar}, this paper devotes itself in proposing an efficient polar-coded forward error correction for multi-level cell (MLC) NAND flash memory.

\subsection{Contributions}
To balance the performance and delay, this paper proposes a pre-check scheme based on polar code for MLC NAND flash. Our main contributions are notably:

\begin{itemize}
  \item We propose the pre-check scheme to arrange pure-soft, quantized-soft, and binary-input polar decoders in different life-stages of SSD.
  \item We have proved that polar code is a balanced code for which each codeword contains an equal\\
      number of zero and one bits.
  \item We propose a well-designed hard-decision binary-input polar decoder. This decoder directly employs $1$-bit hard results returned from the voltage detector and utilizes a single XOR gate to calculate\\
      log-likelihood ratios (LLRs).
  \item We compare the complexities of binary-input SC polar decoder, SC polar decoder, binary-input bit-flipping LDPC decoder, and layered BP polar decoder. Results show that binary-input SC polar decoder has the lowest complexity given a target error performance. Besides, it also has better performance than
      traditional hard-decision bit-flipping LDPC decoder.
  \item
         \begin{minipage}[!h]{.95\textwidth}
           We propose a new quantized-soft polar decoder with refined boundary-defining scheme to improve the empirical method.
         \end{minipage}
  \item We clarify that \Gray code is the optimal scheme to map $2$ bits in $1$ cell.
\end{itemize}


\subsection{Notations}
Let $L$ and $\mathbb{L}$ designate likelihood ratio (LR) and LLR, respectively. Sets are denoted by uppercase calligraphic letters as $\mathcal{A}$. We indicate the probability density function (PDF) of a voltage distribution $i$ by $p^{(i)}$. The uppercase letter $P$ designates probability cumulated by PDFs. The entropy function is $H$.

\subsection{Paper outline}
The remainder of this paper is organized as follows. Section \ref{sec:model and polar} reviews background of NAND flash and polar codes. Section \ref{sec:scheme} proposes the \Gray mapping scheme and pre-check scheme. Three polar decoders are discussed in this section too. In Section
\ref{sec:Binary decoder design}, hardware architecture of proposed binary-input decoder is detailed. In Section \ref{sec:simulation}, performance and complexity are compared for different decoders. Finally, Section \ref{sec:conclusion} concludes this paper. Proof for \Gray mapping scheme and the correction of previous work \cite{kim2012performance} are presented in Appendix.

\section{Background of MLC NAND flash memory and polar codes}\label{sec:model and polar}
\subsection{Modeling of NAND flash memory}\label{subsec:model}
Floating gate transistors constitute the NAND flash memory \cite{li2010improving}. Programming is an operation which stepwise injects a certain quantity of charges to achieve a target voltage. Unavoidably influenced by multiple interferences, voltages will turn into wide ranges, which results in overlapped regions.

The voltage distribution adopted in this work originates from \cite{Intel97}. \Gaussian distribution is selected for both convenience and accuracy of modeling \cite{cai2012error}.

For design purposes, each cell is initialized with $4$ distributions away from each other. However, these distributions gets closer with increasing program/erase (P/E) cycles and multiple interferences. Raw error happens when overlapped regions exist.

\subsection{Basics of polar codes}\label{sec:Basics of polar}
Proposed by E. Ar\i kan in \cite{Arikan09}, polar codes have the capability of achieving the symmetric capacity $I(W)$ of any given B-DMC $W$, so long as the code length $N$ goes to infinity. To better understand polar codes, LLR-based min-sum SC decoding algorithm \cite{leroux2011hardware} is introduced below.

In an arbitrary code with parameter $(N, K, \mathcal{A}, u_{A^c})$, code length and information length are represented by $N$ and $K$. Source vector, the input vector of SC encoder, is denoted by $u_1^N$, which consists of an information part $u_\mathcal{A}$ and a frozen part $u_{\mathcal{A}^c}$. Note that frozen bits $u_{\mathcal{A}^c}$ are usually set to $0$.

The LLR-based min-sum SC decoding algorithm is defined as
\begin{equation}\label{eq:Type I_PE_LLR}
  \mathbb{L}_N^{2i}(y_1^N,\hat u_1^{2i-1})=(-1)^{\hat u_{2i-1}}\mathbb{L}_{N/2}^{(i)} \left( y_{1}^{N/2},\hat u_{1,o}^{2i-2}\oplus u_{1,e}^{2i-2} \right) + \mathbb{L}_{N/2}^{(i)} \left( y_{N/2+1}^{N},\hat u_{1,e}^{2i-2} \right),
\end{equation}

\begin{equation}\label{eq:Type II_PE_Min Sum}
\begin{aligned}
  \mathbb{L}_N^{2i-1}(y_1^N,\hat u_1^{2i-1})\simeq  &\text{sgn} \left[\mathbb{L}_{N/2}^{(i)}\left( y_{1}^{N/2},\hat u_{1,o}^{2i-2}\oplus u_{1,e}^{2i-2}\right)\right] \text{sgn} \left[ \mathbb{L}_{N/2}^{(i)} \left( y_{N/2+1}^{N},\hat u_{1,e}^{2i-2} \right) \right] \cdot\\ &\text{min} \left[\left|\mathbb{L}_{N/2}^{(i)}\left( y_{1}^{N/2},\hat u_{1,o}^{2i-2}\oplus u_{1,e}^{2i-2}\right)\right|, \left|\mathbb{L}_{N/2}^{(i)} \left( y_{N/2+1}^{N},\hat u_{1,e}^{2i-2} \right)\right| \right].
\end{aligned}
\end{equation}

The symbol $\mathbb{L}$ in (\ref{eq:Type I_PE_LLR}) and (\ref{eq:Type II_PE_Min Sum}) denotes LLR
\begin{equation}\label{eq:defination of LLR}
  \mathbb{L}_N^{i}(y_1^N,\hat u_1^{i-1})\triangleq \ln L_N^{(i)}(y_1^N,\hat u_1^{i-1}),
\end{equation}
where $L_N^{(i)}(y_1^N,\hat u_1^{i-1})$ is the LR and $\hat u_i$ $(i\in \mathcal{A})$ is estimated as
\begin{equation}\label{eq:SC Alg. LLR}
\hat u_i = \left\{ \begin{array}{l}
  0,\quad \text{if } \mathbb{L}_N^{(i)}(y_1^N, \hat u_1^{i-1})\geq0;\\
  1,\quad \text{otherwise.}
\end{array} \right.
\end{equation}

The hardware architecture of this algorithm is explained in \cite{zhang2013low}.

\section{Multi-strategy ECC scheme}\label{sec:scheme}
In this section, we first demonstrate the adopted \gray mapping scheme. Then we propose the pre-check scheme with multi-strategy ECC and $3$ corresponding polar decoders.

\subsection{Gray mapping and detection}\label{subsec:Gray}

\begin{figure}[!t]
\begin{minipage}[t]{0.55\linewidth}
  \centering
  \includegraphics[width=.9\linewidth]{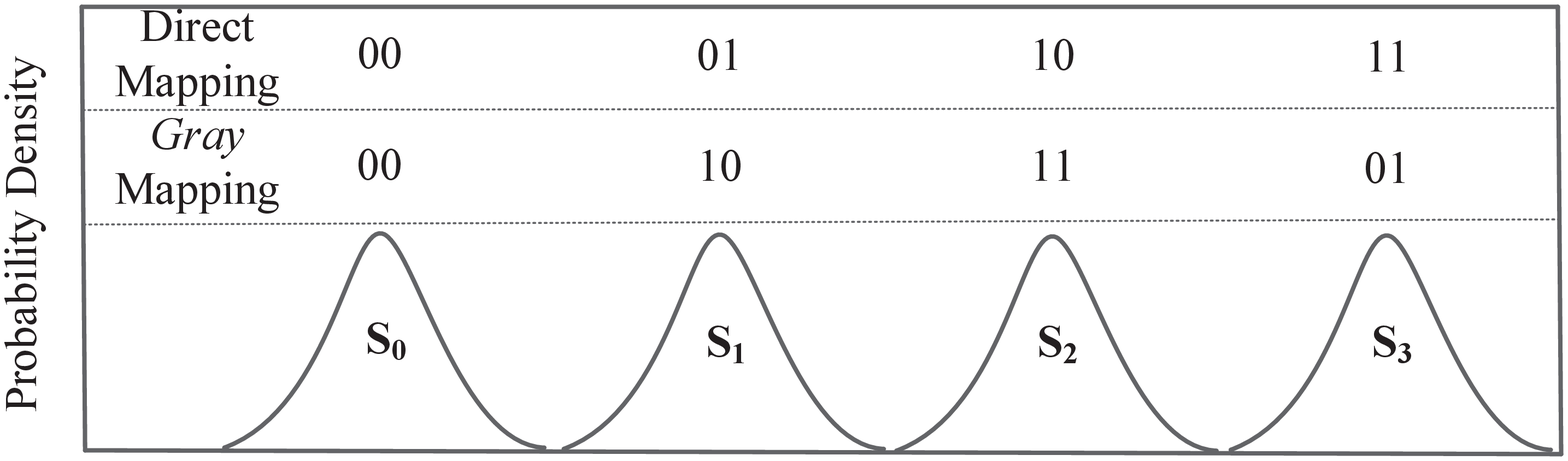}\\
  \caption{Modeling and \Gray mapping scheme.}\label{fig:c state detailed}
\end{minipage}%
\begin{minipage}[t]{0.5\linewidth}
\centering
\includegraphics[width=0.9\linewidth]{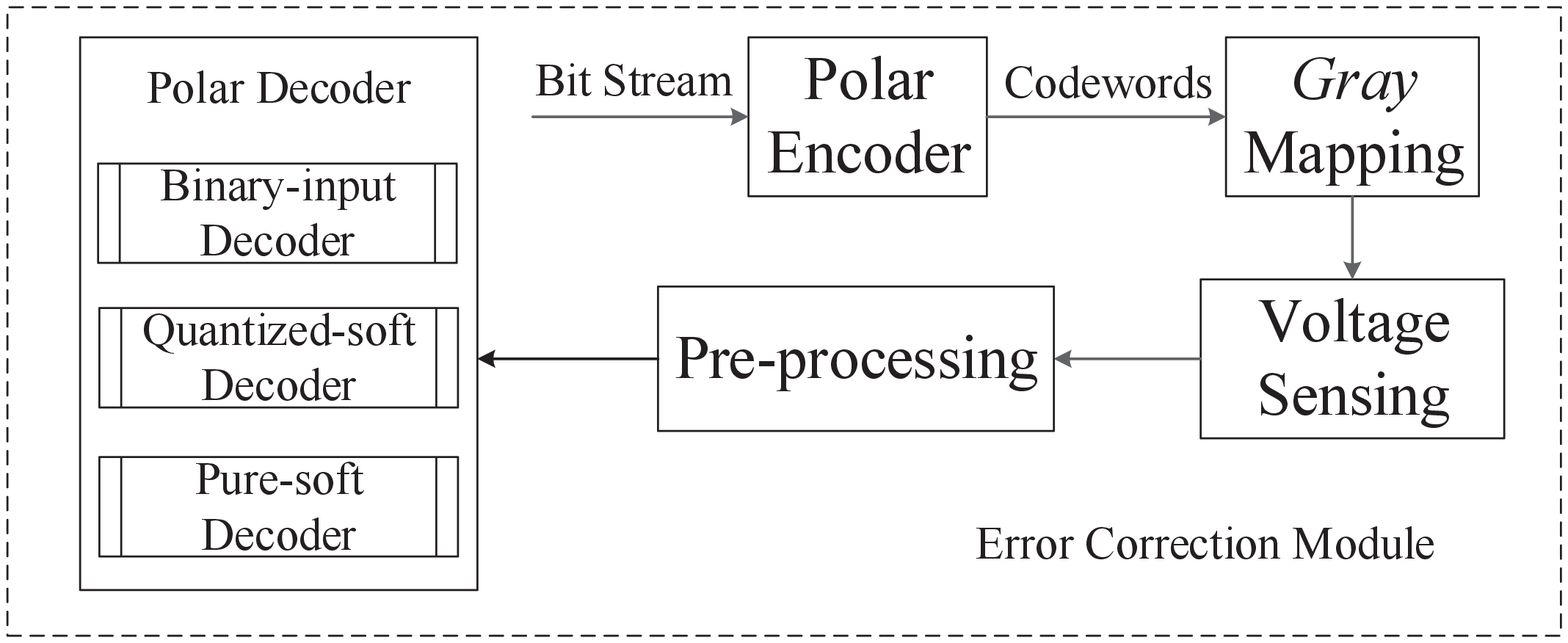}\\
  \caption{Error correction module in SSD controller.}\label{fig:fun block}
\end{minipage}%
\end{figure}

The programmed symbols for each state of MLC NAND flash memory are shown in Figure~\ref{fig:c state detailed}. Note that a raw error happens when a state is mistakenly considered for its neighboring states. Moreover, $S_1$ and $S_2$ have $2$-bit difference under direct mapping. Hence, we should consider a mapping scheme that is capable of reducing raw error bits. To this end, \gray mapping with minimum difference between adjacent states is the optimal choice. The proof is shown in \ref{appd:proof for gray code}.


\subsection{Control system}\label{subsec:ctrl archi}

\begin{minipage}[!h]{1\textwidth}
The overall architecture of error correction module is illustrated in Figure~\ref{fig:fun block}. Polar decoder will encode the external bit stream into binary codewords. Then these codewords will be pairwise mapped to a certain voltage in each cell. To recover the stored data, the detector first senses a cell several times and compare the stored voltage to reference voltages. After that, pre-check scheme will determine which decoder should be picked and then process the comparison results to LLRs to feed corresponding decoders.
\end{minipage}

\begin{figure}[!t]
\begin{minipage}[t]{0.55\linewidth}
  \centering
  \includegraphics[width=.85\textwidth]{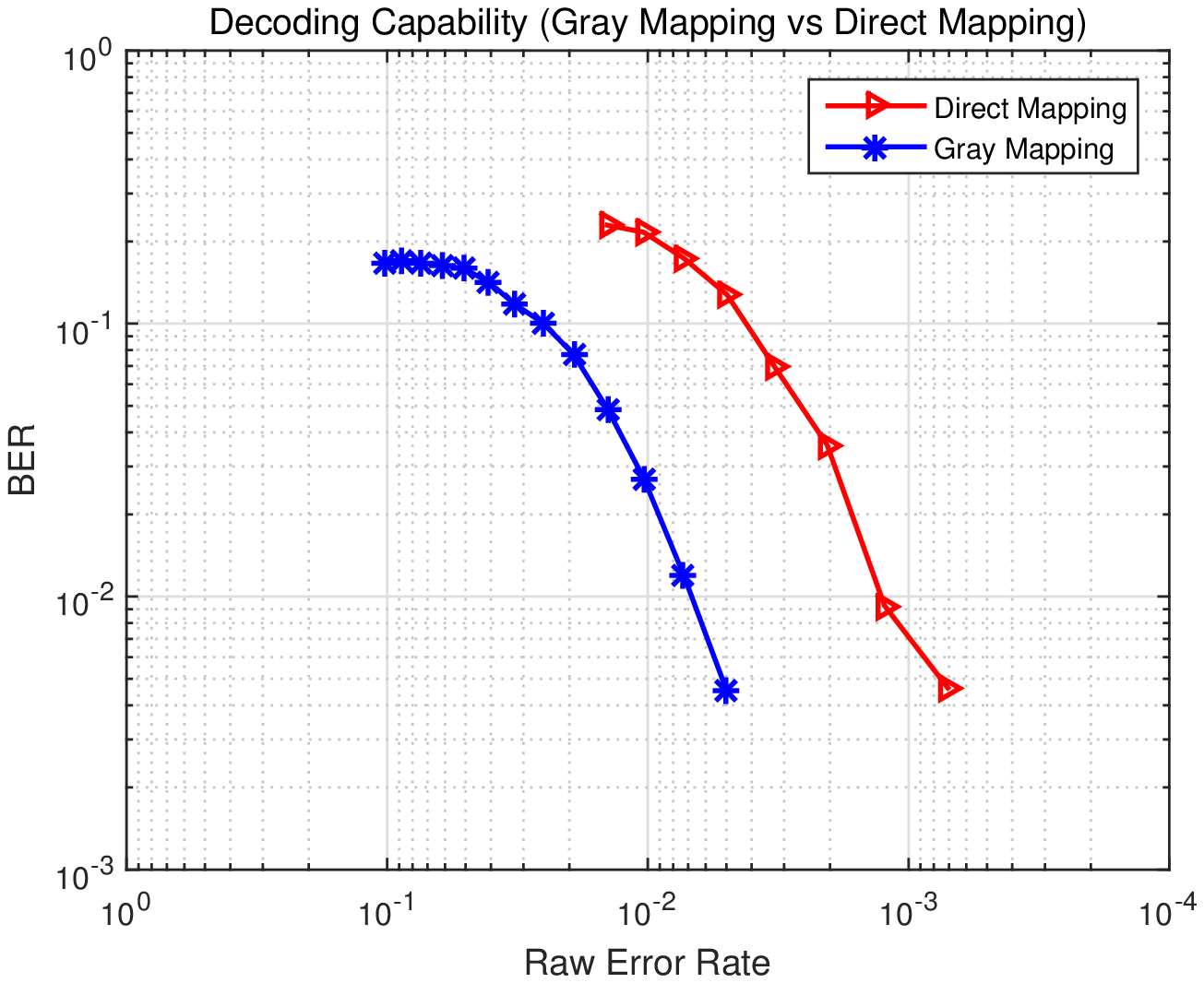}\\
  \caption{
  BER results of a (1024,512) hard-decision polar decoder}
  \label{fig:GrayGain}
\end{minipage}%
\begin{minipage}[t]{0.55\linewidth}
\centering
  \centering
  \includegraphics[width=0.85\linewidth]{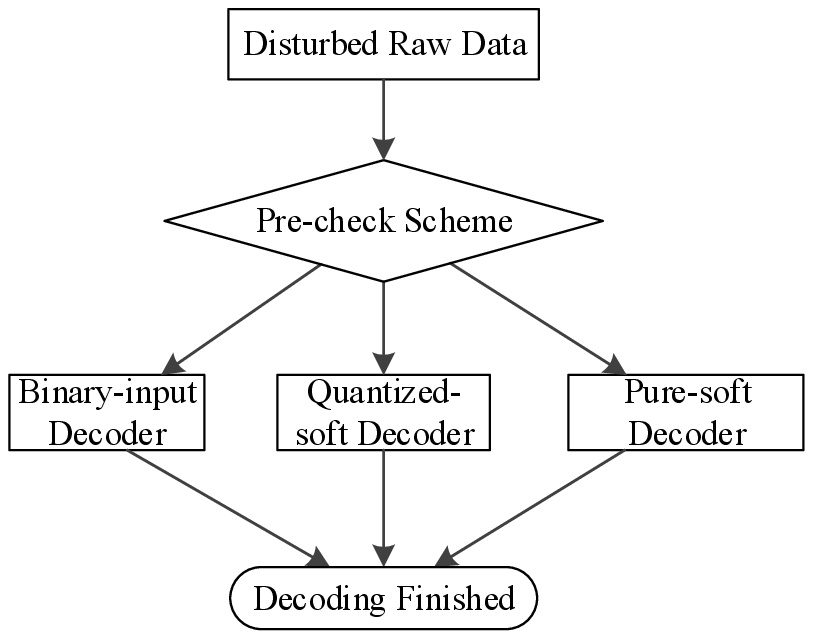}\\
  \caption{Flow chart of pre-check scheme.}\label{fig:flow control}
\end{minipage}
\end{figure}

Figure~\ref{fig:flow control} illustrates the flow of each step in the pre-check scheme. Cell state will be checked at the beginning to determine which decoder should be picked. When cell distortion appears slight, the binary-input decoder is chosen owing to its low decoding latency. When distortion is getting worse, soft-decision decoders should be selected to guarantee data integrity.

\subsection{Pre-check scheme}\label{subsec:pre-check}
This scheme aims to select an optimal decoder in accordance of the condition to meet the demand for storage reliability.

Assume the mean values of four states are $V$, $2V$, $3V$, and $4V$ respectively and the standard deviation is $\sigma$,  which is identical for all distributions.

The cell state can be expressed as a set of equations as
\begin{equation} \label{eq:cell state}
{p^{(i)}}(x) = \dfrac{1}{{\sqrt {2\pi } \sigma }}\exp ( - \dfrac{{{{(x - (i+1) V)}^2}}}{{2{\sigma ^2}}}) \qquad (i = 0,1,2,3).
\end{equation}
By solving (\ref{eq: HARD boundaries})
\begin{equation} \label{eq: HARD boundaries}
p^{(i)}(x) = {p^{(i+1)}}(x) \qquad (i = 0,1,2),
\end{equation}
we can obtain intersections $[R_1,R_2,R_3]$ between $4$ distributions which are
\[
R_1 = \frac{3}{2} V, \quad R_2 = \frac{5}{2} V ,\quad R_3 = \frac{7}{2} V.
\]
Since mean values are uniformly distributed and standard deviations are identical, reference voltage $R_i$ is the mid-value between $\mu^{(i-1)}$ and $\mu^{(i)}$. A raw error will occur when the sensed voltage gets across the reference voltage. For example, if a voltage of state $p^{(0)}$ is greater than $R_1$, it is more likely to be considered as a voltage in $p^{(1)}$ (i.e., an error happens). Therefore, we can calculate the raw error probability for each overlapped region by
\begin{equation}\label{eq:raw error probability}
  {P_E}=
  \begin{cases}
    &\int_{ - \infty }^{{R_i}} {{p^{(i)}}(x)dx}  = 0.1995\sqrt {2\pi }[1 - erf(\frac{\sqrt {2V}}{4\sigma})]\quad \text{leftmost and rightmost distributions};\\
    &\int_{ - \infty }^{{R_i}} {{2p^{(i)}}(x)dx}  = 0.3990\sqrt {2\pi }[1 - erf(\frac{\sqrt {2V}}{4\sigma})]\quad \text{middle distributions}.
  \end{cases}
\end{equation}

\begin{minipage}[!h]{1\linewidth}
  $P_E$ is a function of variable $\frac {\sqrt{V}}{\sigma}$, where $V$ is the distance between two adjacent distributions and $\sigma$ is the standard deviation. In NAND flash memory, the values of $\sqrt{V}$ and $\sigma$ change over time due to voltage shifting and cell distortion. Since $P_E$ is monotonically decreasing with $\frac {\sqrt{V}}{\sigma}$ and $\frac {\sqrt{V}}{\sigma}$ is decreasing over time (the experiment in \cite{cai2012error} has shown that the signal-to-noise ratio (SNR) in the NAND flash memory degrades about $0.13$dB per $1$k P/E cycles), the value of $P_E$ is increasing.
\end{minipage}

With numerical $P_E$, we can set several thresholds to adjust the decoding scheme to satisfy performance requirements of the system.

\subsection{Pure-soft decoder}\label{subsec:pure soft}
The sensed voltage needs to be converted into digital LLR to feed the pure-soft decoder.

Given the model of NAND flash memory in Section \ref{sec:model and polar}, the whole voltage range can be described with $4$ \Gaussian distributions indicated by $p^{(0)}(x),\;p^{(1)}(x),\;p^{(2)}(x)$, and $p^{(3)}(x)$. To obtain the definition of LLR in NAND flash memory, there are some basic ideas that need to be clarified.

  \lemma[]\label{Lemma:Probability of 0&1}

\begin{minipage}[!h]{.8\linewidth}
Polar code is a balanced code  for which each codeword contains an equal number of zero and one bits.\end{minipage}

\proof
  The codeword $x_1^N$ and the $i$-th element of $x_N^{(i)}$ are constructed as
  \[
    x_1^N = u_1^N\cdot G_N,\qquad x_N^{(i)} = u_1^NG_N^{(i)},
  \]
  where $u_1^N$ is the source information, $G_N$ is the generator matrix and $G_N^{(i)}$ denotes the $i$-th column of $G_N$.

  With the property of multiplication in $GF(2)$, whether $x_N^{(i)}$ is $0$ or $1$ is only determined by the number of $1$'s in $u_1^N$ whose corresponding places in $G_N^{(i)}$ are $1$. For example, if $N=4,i=2$, then we have
  \begin{equation}\label{eq:proof}
    x_4^{(2)} = u_1^4\cdot [0\;0\;1\;1]^T = u_3+u_4.
  \end{equation}
  Since some elements in $G_N$ are $0$, only a part of elements in $u_1^N$ participate in the calculation. In the example of (\ref{eq:proof}), only $u_3$ and $u_4$ are concerned.

  Assume that the number of $1$'s in $G_N^{(i)}$ is $G_i$. The probability for $x_N^{(i)}$ being $0$ or $1$ can be denoted by
  \[
  \small
  \begin{aligned}
    P(x_N^{(i)}=0) = C_{G_i}^0(\frac12)^0(\frac12)^{G_i-0}+C_{G_i}^2(\frac12)^2(\frac12)^{G_i-2}+\ldots;\;P(x_N^{(i)}=1) = C_{G_i}^1(\frac12)^1(\frac12)^{G_i-1}+C_{G_i}^3(\frac12)^2(\frac12)^{G_i-3}+\ldots\\
  \end{aligned}
  \]

  \[
  \begin{aligned}
    \because\; & C_n^0+C_n^2+\ldots = C_n^1+C_n^3+\ldots=2^{n-1}\therefore P(x_N^{(i)}=0) = P(x_N^{(i)}=1) =\frac12.
  \end{aligned}
  \]

 \begin{minipage}[!h]{1\textwidth}
     Lemma \ref{Lemma:Probability of 0&1} is the foundation for LLR calculation in NAND flash memory. This \emph{a priori}  property guarantees the usage of \emph{Bayes} Law within LLR calculation for all the polar decoders discussed in this paper.
 \end{minipage}

\lemma[]\label{Lemma:Calculation of Soft LLR}
{For any stored bit $b_i$, its LLR is defined as
  \begin{equation}\label{eq:LLR_General}
  \begin{aligned}
    \mathbb{L}({b_i}) = \log \frac{{\sum\limits_{k \in {O_i}} {{p^{(k)}}({V_d})} }}{{\sum\limits_{k \in {Z_i}} {{p^{(k)}}({V_d})} }},
  \end{aligned}
\end{equation}
where $p^{(k)}$ denotes the $k$-th PDF of voltage distribution, $V_d$ denotes the sensed voltage, $O_i$ contains distributions where $b_i = 1$ and $Z_i$ contains distributions where $b_i = 0$.}

\begin{figure}[!t]
 \begin{minipage}[t]{0.5\linewidth}
  \centering
  \includegraphics[width=.99\textwidth]{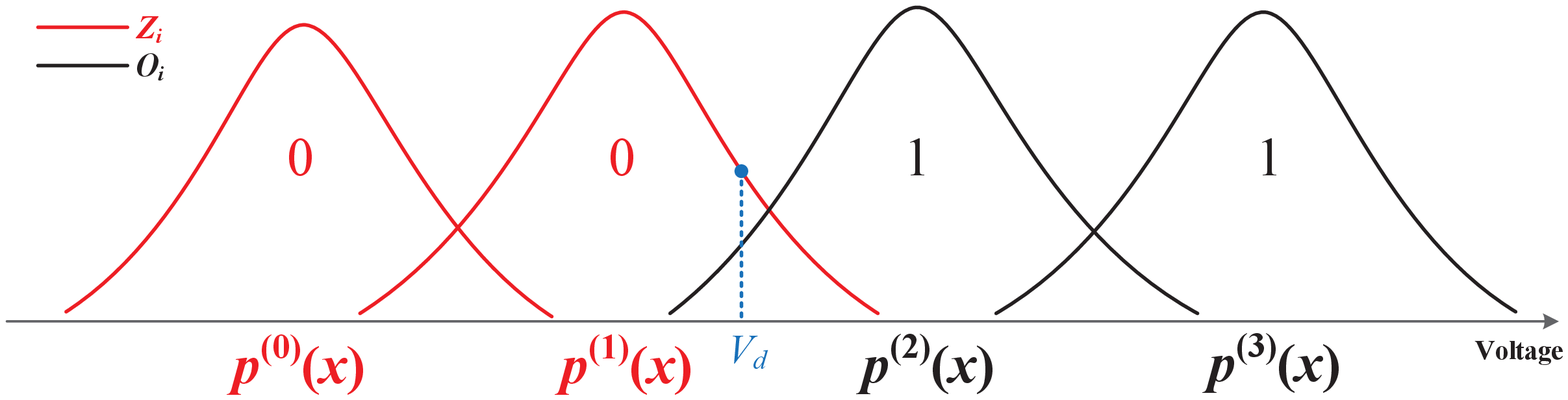}
  \caption{An example for soft LLR calculation based on a specific division of $O_i$ and $Z_i$. }\label{fig:Cell_state_double_color}
 \end{minipage}
 \begin{minipage}[t]{0.5\linewidth}
  \centering
  \includegraphics[width=.8\textwidth]{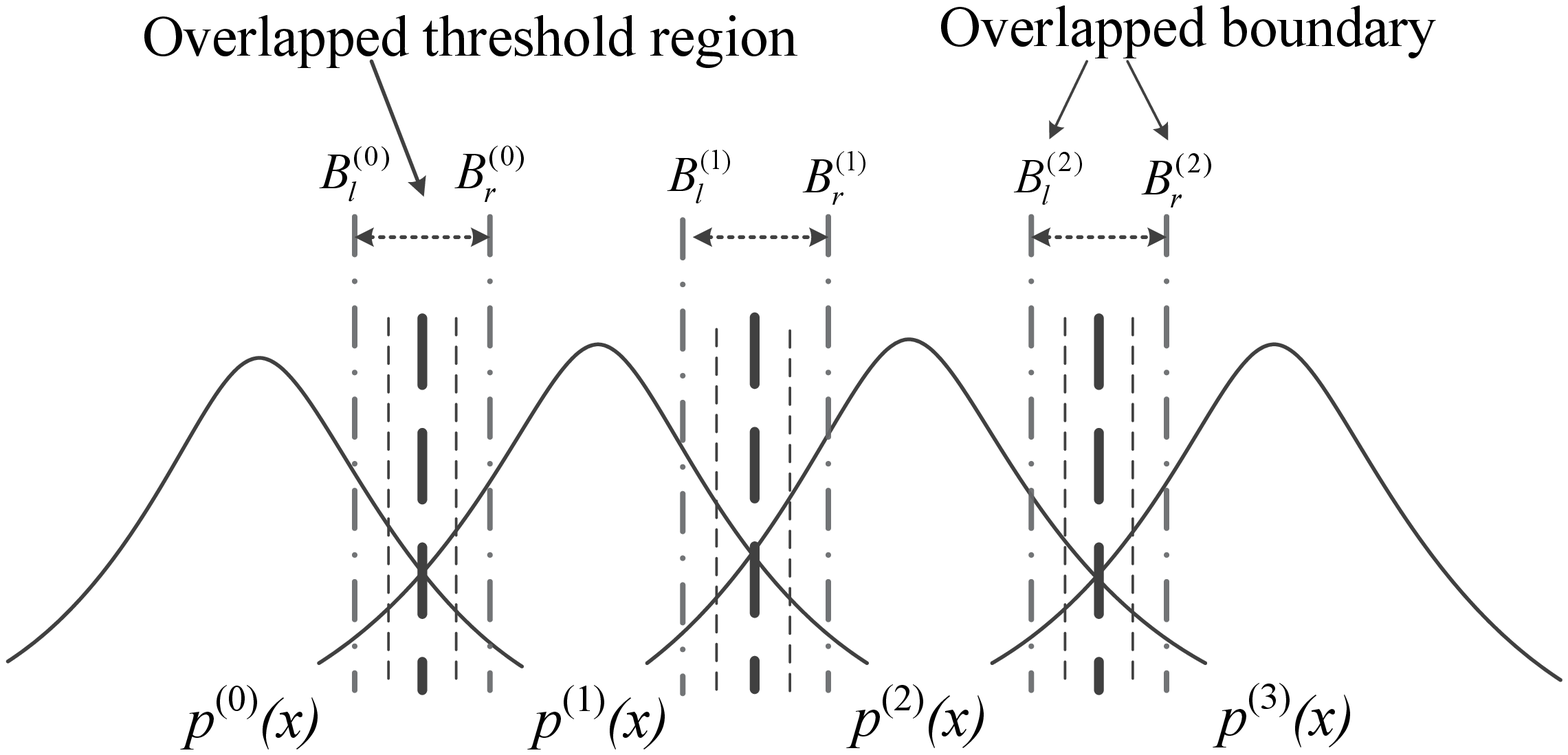}
  \caption{Non-uniform sensing operations \cite{kim2012performance}.}\label{fig:c state Q-soft}
 \end{minipage}
\end{figure}

\proof
  According to the definition, the LLR of $b_i$ should be denoted by
  \begin{equation}\label{eq:L_bi_poste}
    \begin{aligned}
    \mathbb{L}({b_i}) = \log \frac{{p({b_i} = 1|{V_d})}}{{p({b_i} = 0|{V_d})}}.
    \end{aligned}
  \end{equation}
However, considering the difficulty of directly acquiring the \textit{a posteriori} probability $p(b_i|V_d)$, it is simple to transform (\ref{eq:L_bi_poste}) into the form of likelihood function according to the \emph{Bayes} theorem as
\begin{equation}\label{eq:L_bi}
\begin{aligned}
    \mathbb{L}({b_i}) = \log \frac{{p({b_i} = 1|{V_d})}}{{p({b_i} = 0|{V_d})}} = \log \frac{{p({V_d}|{b_i} = 1)}p(b_i = 1)}{{p({V_d}|{b_i} = 0)}p(b_i = 0)} = \log \frac{{p({V_d}|{b_i} = 1)}}{{p({V_d}|{b_i} = 0)}},
\end{aligned}
\end{equation}
where $\frac {p(b_i = 1)}{p(b_i = 0)}=1$, according to Lemma \ref{Lemma:Probability of 0&1}, and $p(V_d|b_i)$ is the summation of PDFs when $b_i$ is settled. Therefore, the LLR of $b_i$ is exactly the form in (\ref{eq:LLR_General}). Note that $O_i$ and $Z_i$ are different according to the adopted mapping scheme. An example is shown in Figure~\ref{fig:Cell_state_double_color}.

\subsection{Quantized-soft decoder}\label{subsec:q-soft}
The LLR calculation mentioned in Section~\ref{subsec:pure soft} can achieve the best performance of error correction. However, it requires an accurate value of the sensed voltage, which is unrealistic in circuits. Therefore, a proper scheme which can balance the numerical accuracy and sensing latency is highly needed.

\subsubsection{Problems in quantized-soft decoder}
The main constrain is that the detector can only return a comparison result between the sensed voltage and pre-set references which we call ``hard result", containing only $1$-bit information.

This raises two problems. The first one is obtaining proper references (or boundaries). The definition of overlapped regions is crucial to calculate LLRs.

Another problem is the number of sensing operations. Considering that LLR contains information more than $1$ bit, we need multiple sensing operations to convert hard results into LLR. An example is shown in Figure~\ref{fig:c state Q-soft} \cite{kim2012performance}.

\subsubsection{Boundaries defined by constant ratio}\label{subsec:constant ratio}
In our previous work \cite{song2016polar}, we adopted the boundary-defining scheme that was proposed in \cite{dong2011use} and expanded in \cite{kim2012performance}. In this section, we show the basic idea in \cite{dong2011use} and the re-derived quadratic equation set which differs from the equations in \cite{kim2012performance}.

$B_l^{(k)}$ and $B_r^{(k)}$ are $2$ boundaries restricting the $k$th region, and $R$ is a pre-settled ratio. The relation among $B_l^{(k)}$, $B_r^{(k)}$ and $R$ is as
\begin{equation}\label{eq:Bond.1}
\frac{{{p^{(k)}}({B_l}^{(k)})}}{{{p^{(k + 1)}}({B_l}^{(k)})}} = \frac{{{p^{(k + 1)}}({B_r}^{(k)})}}{{{p^{(k)}}({B_r}^{(k)})}} = R,
\end{equation}
where $p^{(k)}(x)$ is the $k$th voltage distribution. Under \Gaussian estimation, this calculation is significantly simplified compare with \cite{dong2011use}.

Let $\sigma^2 _k$ and $\mu _k$ be the variation and mean value of $p^{(k)}$, then we have
\begin{equation}\label{eq:Bond}
\begin{aligned}
 \left\{
 \begin{array}{l}
    2\boldsymbol{\sigma _k^2\sigma _{k + 1}^2}\log (\dfrac{{{\sigma _k}}}{{{\sigma _{k + 1}}}}R) =  - \sigma _{k + 1}^2{({B_l}^{(k)} - {\mu _k})^2}+ \sigma _k^2{({B_l}^{(k)} - {\mu _{k + 1}})^2},\\
    2\boldsymbol{\sigma _k^2\sigma _{k + 1}^2}\log (\dfrac{{{\sigma _{k + 1}}}}{{{\sigma _k}}}R) =  - \sigma _k^2{({B_r}^{(k)} - {\mu _{k + 1}})^2}+ \sigma _{k + 1}^2{({B_r}^{(k)} - {\mu _k})^2}.
 \end{array} \right.
\end{aligned}
\end{equation}
Eq.~(\ref{eq:Bond}) is derived from (\ref{eq:Bond.1}). The work of \cite{kim2012performance} does not show the derivation, whereas their result is slightly wrong. We add red corrections and further derive it in \ref{appd:derivation}.

\subsubsection{Boundaries defined by stepwise mutual information}
The boundary-defining scheme of constant ratio mentioned in Section \ref{subsec:constant ratio} is effective to locate the overlapped regions. However, there still remains an unsolved problem that the value of $R$ is mostly determined by empirical evidence.

A different scheme called maximum mutual information (MMI) is proposed in \cite{wang2011soft} which aims to set quantization boundaries that maximize the mutual information. MMI quantizes the whole voltage range into $(M+1)$ regions with $M$ sensing operations.

However, MMI is a general case instead of an optimal choice for boundary selection because the mutual information defined in \cite{wang2011soft} is calculated for each region instead of original bits, whereas LLRs are calculated bitwise. In this work, we calculate mutual information for the most significant bit (MSB) and the least significant bit (LSB) separately, which we call stepwise mutual information (SMMI).

\begin{figure}[!t]
 \begin{minipage}[t]{0.3\linewidth}
    \centering
    \includegraphics[width=0.9\linewidth]{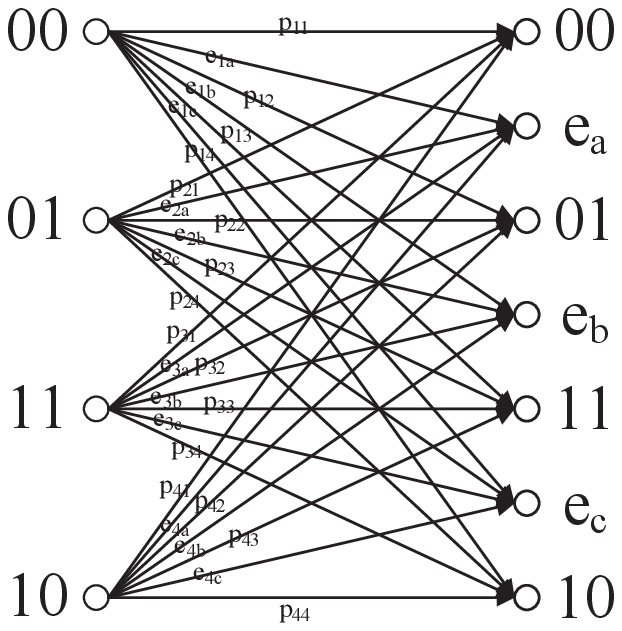}
    \caption{$4$-input, $7$ output\\MLC model for MMI scheme.}
    \label{fig:Entrophy_MMI}
 \end{minipage}
 \begin{minipage}[t]{0.7\linewidth}
    \centering
    \includegraphics[width=0.9\linewidth]{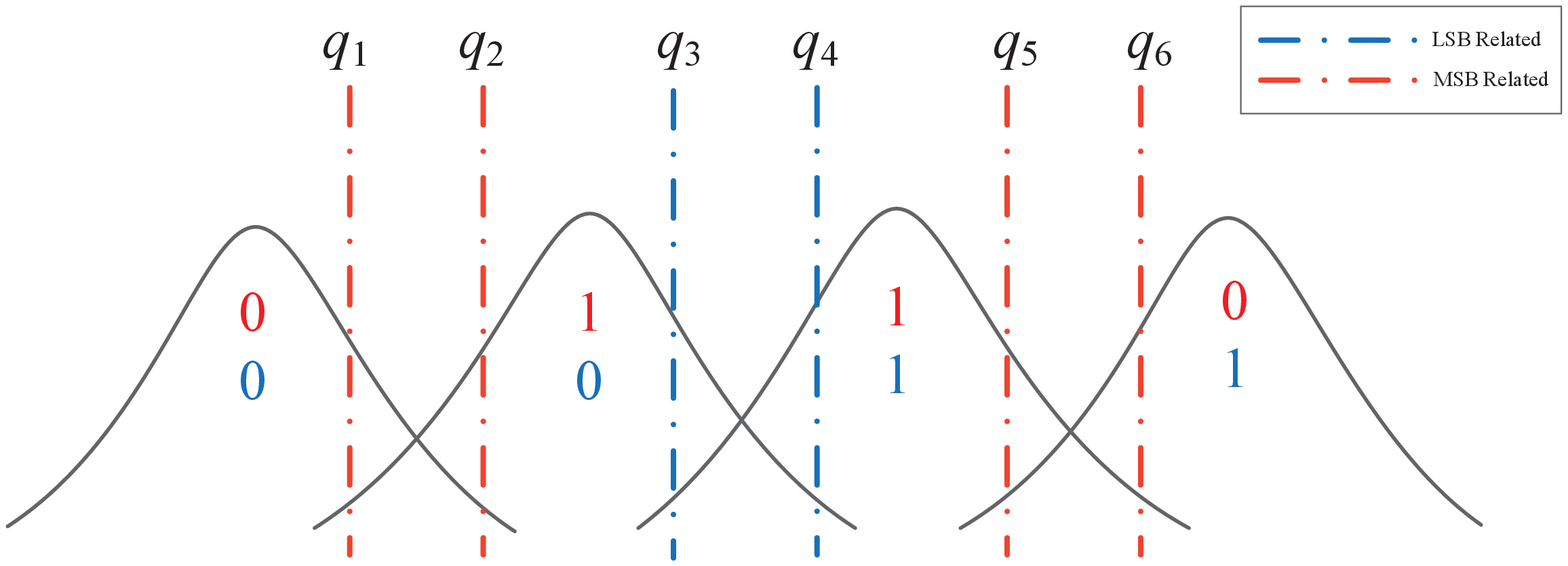}
    \caption{Different references for the LSB and the MSB.}
    \label{fig:Entrophy_SMMI}
 \end{minipage}
\end{figure}

Figure~\ref{fig:Entrophy_SMMI} shows the relationship between reference voltages and mapped bits. It is obvious that the judgement of the LSB only relates to $2$ quantization boundaries $q_3$ and $q_4$. Similarly, $q_1,q_2, q_5$, and $q_6$ are responsible for sensing operation of the MSB. We take the LSB as an example to demonstrate the channel and the entropy calculation under SMMI strategy.

\begin{figure}[!t]
 \begin{minipage}[t]{0.6\linewidth}
    \centering
    \includegraphics[width=0.95\linewidth]{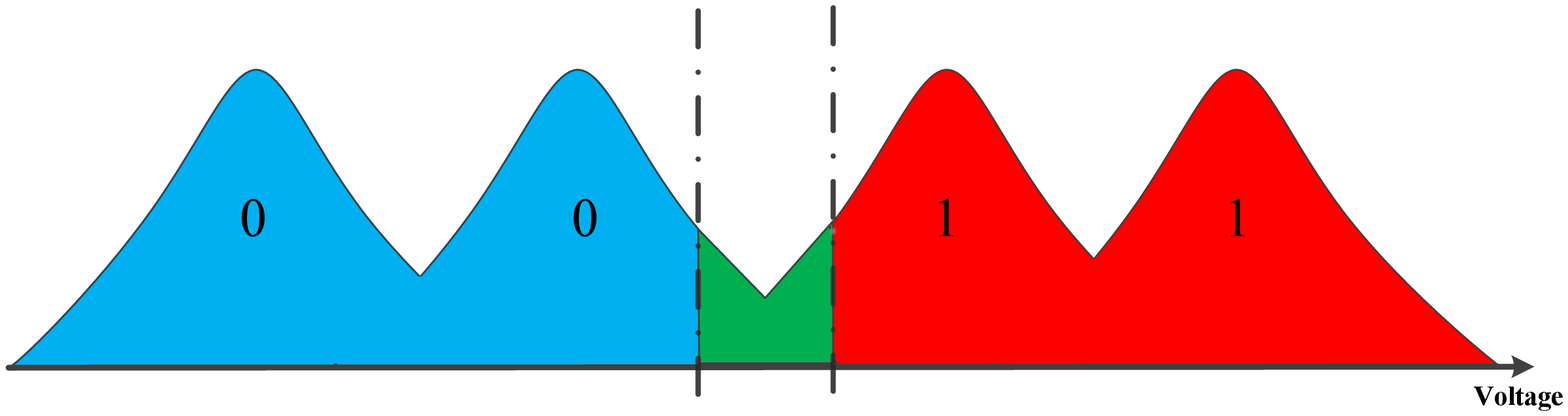}
    \caption{Quantization boundaries for the LSB in MLC.}
    \label{fig:Entrophy_LSB}
 \end{minipage}%
 \begin{minipage}[t]{0.4\linewidth}
    \centering
    \includegraphics[width=0.5\linewidth]{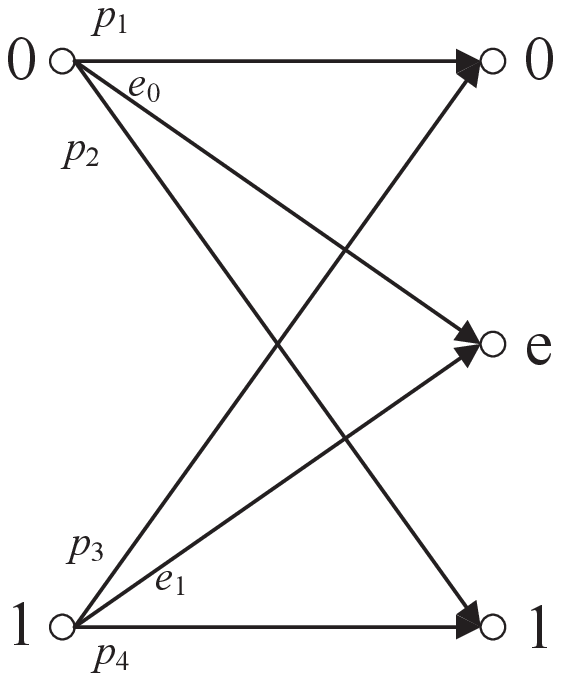}
    \caption{Channel model for the LSB.}
    \label{fig:MLC_MODEL_LSB}
 \end{minipage}
\end{figure}

In Figure~\ref{fig:Entrophy_LSB}, the whole range is separated into $3$ quantized regions, hence this quantization model is equivalent to a $2$-input, $3$-output channel model with $X\in\{0,1\}$ and $Y\in\{0,e,1\}$ given in Figure~\ref{fig:MLC_MODEL_LSB}, which is similar to the model of single-level cell (SLC) NAND flash memory with $2$ reads in \cite{wjd2012mutual}.

According to Lemma \ref{Lemma:Probability of 0&1}, $X$ sends $0$ and $1$ under equal probability. Therefore, the mutual information $I$ between $X$ and $Y$ is calculated as
\begin{equation}\label{eq:MLC_MUTUAL_LSB}
\begin{aligned}
    I(X;Y)=H(Y)-H(Y|X)=H(\frac{p_{1}+p_{3}}{2},\frac{e_{0}+e_{1}}{2},\frac{p_{2}+p_{4}}{2})-\frac12 H(p_1,e_0,p_2)-\frac12 H(p_3,e_1,p_4).
\end{aligned}
\end{equation}

For a settled voltage distribution, the mutual information between $X$ and $Y$ can be numerically maximized to obtain desired boundaries $q_3$ and $q_4$ that yield the SMMI.

\subsubsection{Practical SMMI boundary calculation}
\begin{figure}[!t]
    \centering
    \includegraphics[width=0.6\linewidth]{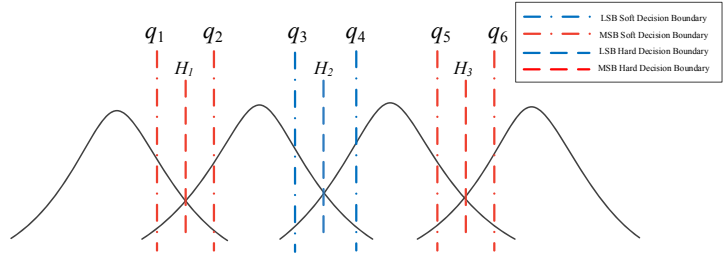}
    \caption{$9$ reference in practical scheme.}
    \label{fig:Entrophy_SMMI_practical}
\end{figure}

\begin{figure}[!t]
 \begin{minipage}[t]{0.5\linewidth}
    \centering
    \includegraphics[width =.45\linewidth]{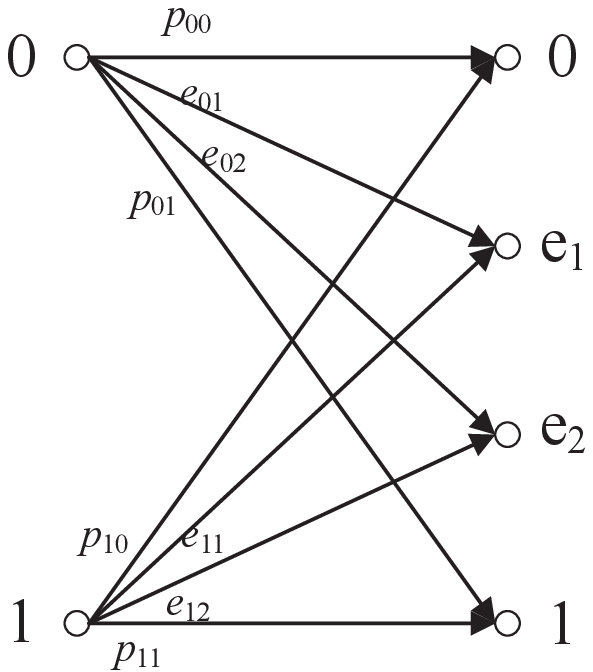}
    \caption{Channel models for the LSB in practical scheme.}
    \label{fig:LSB_MODEL4}
 \end{minipage}%
 \begin{minipage}[t]{0.5\linewidth}
    \centering
    \includegraphics[width=.45\linewidth]{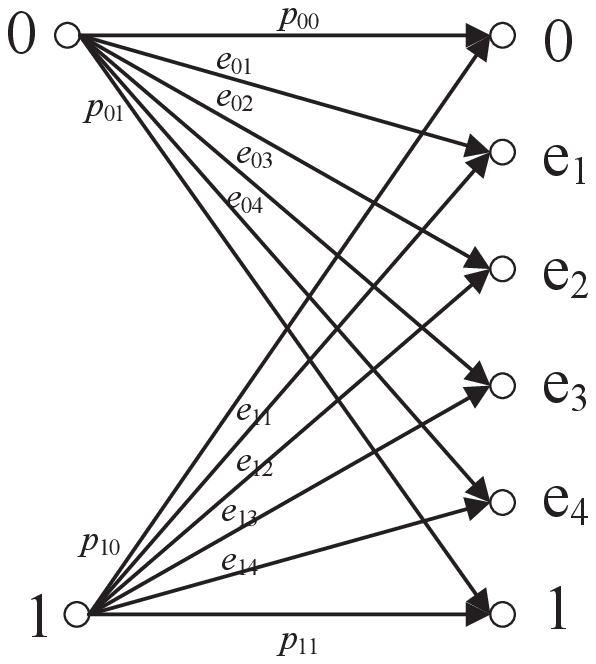}
    \caption{Channel models for the MSB in practical scheme.}
    \label{fig:MSB_MODEL6}
 \end{minipage}
\end{figure}

In the MMI example shown above, a $4$ input, $7$ output MLC model shown in Figure~\ref{fig:Entrophy_MMI} was adopted for illustration purposes. However, there are at least $3$ sensing operations in $1$ overlapped region in a practical control system as demonstrated in Figure~\ref{fig:Entrophy_SMMI_practical}, where the intersections of two distributions in the middle are called ``hard-decision boundaries'' and $\{q_i, i = 1,2,...6\}$ mentioned before are called ``soft-decision boundaries". Channel models for the LSB and the MSB in this scheme are shown in Figure~\ref{fig:LSB_MODEL4} and Figure~\ref{fig:MSB_MODEL6}.

The mutual information for the LSB in this case is calculated as
\begin{equation}\label{eq:MLC_MUTUAL_Prac_LSB}
\small
\begin{aligned}
    I(X;Y)&=H(Y)-H(Y|X)\\
    &=H(\frac{p_{00}+p_{10}}{2},\frac{e_{01}+e_{11}}{2},\frac{e_{02}+e_{12}}{2},\frac{p_{01}+p_{11}}{2})-\frac12 H(p_{00},e_{01},e_{02},p_{01})-\frac12 H(p_{10},e_{11},e_{12},p_{11}),
\end{aligned}
\end{equation}
and the mutual information for the MSB is calculated as
\begin{equation}\label{eq:MLC_MUTUAL_Prac_MSB}
\begin{aligned}
    I(X;Y)=&H(Y)-H(Y|X)
     \\=&H(\frac{p_{00}+p_{10}}{2},\frac{e_{01}+e_{11}}{2},\frac{e_{02}+e_{12}}{2},\frac{e_{03}+e_{13}}{2},\frac{e_{04}+e_{14}}{2},\frac{p_{01}+p_{11}}{2})\\
     &-\frac12 H(p_{00},e_{01},e_{02},e_{03},e_{04},p_{01})-\frac12 H(p_{10},e_{11},e_{12},e_{13},e_{14},p_{11}).
\end{aligned}
\end{equation}

\subsubsection{LLR calculation}
According to (\ref{eq:LLR_General}), quantized LLRs are calculated as follows:
\begin{equation}\label{eq:LLR}
    \mathbb{L}_i^{\textrm{LSB}} = \log \frac{{\int_{{R_i}} {{p^{(2)}}(x)+{p^{(3)}}(x)dx}}} {{\int_{{R_i}} {{p^{(0)}}(x)+{p^{(1)}}(x)dx}}},\quad
    \mathbb{L}_i^{\textrm{MSB}} = \log \frac{{\int_{{R_i}} {{p^{(1)}}(x)+{p^{(2)}}(x)dx}}} {{\int_{{R_i}} {{p^{(0)}}(x)+{p^{(3)}}(x)dx}}}.
\end{equation}

$\mathbb{L}_i^{\textrm{LSB}}$ and $\mathbb{L}_i^{\textrm{MSB}}$ designate LLRs of the LSB and the MSB of the quantization region $R_i$. We take the LSB as an example to further explain (\ref{eq:LLR}).

Under \gray mapping scheme in Section \ref{subsec:Gray} (illustrated in Figure~\ref{fig:c state hard modified}), $p^{(2)}(x)$ and $p^{(3)}(x)$ are $2$ distributions where LSB$=1$. Meanwhile, $p^{(0)}(x)$ and $p^{(1)}(x)$ are distributions where LSB$=0$. Under this condition, the numerator in (\ref{eq:LLR}) which contains the integral with respect to $x$ of PDF $(p^{(2)}(x)+p^{(3)}(x))$ over the interval $R_i$ represents the probability for LSB$=1$. In this way, the denominator is the probability where LSB$=0$.

Under \Gaussian estimation, $Q$-function can easily calculate desired LLRs as
\begin{equation}\label{eq:PE_LLR}
    \mathbb{L}_i^{\textrm{LSB}} = \log \frac{\sum\limits_{j=2,3} Q(\frac{q_r-\mu_j}{\sigma_j})-Q(\frac{q_l-\mu_j}{\sigma_j})}{\sum\limits_{k=0,1} Q(\frac{q_r-\mu_k}{\sigma_k})-Q(\frac{q_l-\mu_k}{\sigma_k})},\quad
    \mathbb{L}_i^{\textrm{MSB}} = \log \frac{\sum\limits_{j=1,2} Q(\frac{q_r-\mu_j}{\sigma_j})-Q(\frac{q_l-\mu_j}{\sigma_j})}{\sum\limits_{k=0,3} Q(\frac{q_r-\mu_k}{\sigma_k})-Q(\frac{q_l-\mu_k}{\sigma_k})}.
\end{equation}

\subsection{Binary-input decoder}\label{subsec:Binary decoder}
\begin{figure}[!t]
  \centering
  \includegraphics[width=0.5\textwidth]{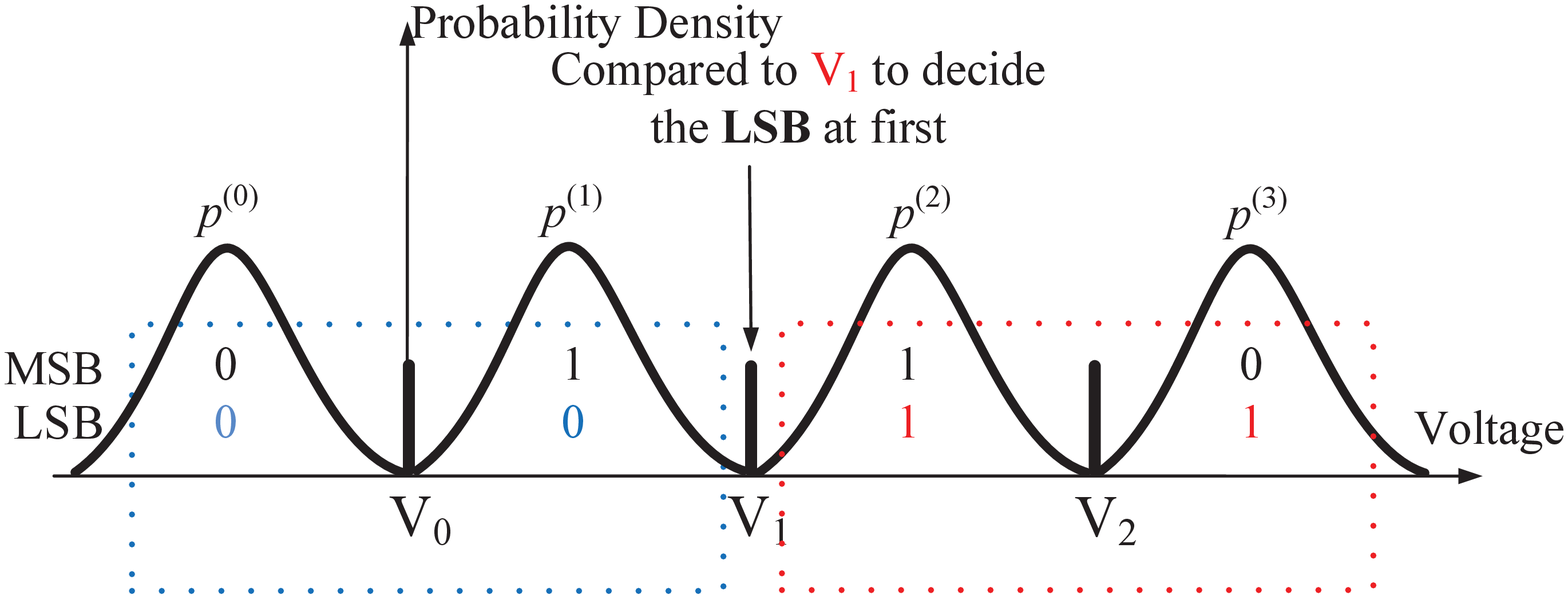}
  \caption{Detection in binary-input decoder.}\label{fig:c state hard modified}
\end{figure}

A sensing strategy is shown in Figure~\ref{fig:c state hard modified}. Three reference voltages are denoted by $V_0$, $V_1$, and $V_2$ which separate $4$ voltage distributions. The detector first compare current voltage with $V_1$ to decide the LSB and then with $V_0$ or $V_2$ to decide the MSB. Detailed description can be found in \cite{song2016polar}.

According to (\ref{eq:SC Alg. LLR}), $\hat u$ is judged by the sign bit of LLR. Therefore, hard results can be fully utilized since they can represent the sign bit of LLR. In other words, they can be transformed into a special form of quantized LLR consisting of only a sign bit, for which it is called ``binary-input decoder".

Magnitude of LLR is not concerned in this scenario and only sign bits will participate in the subsequent calculation, which makes it possible to apply simple bit operations in hardware without adder-subtractors in traditional processing element (PE) design \cite{zhang2012reduced}. This design is hardware-friendly and will be further discussed in Section \ref{sec:Binary decoder design}.

\section{Architecture of proposed binary-input decoders}\label{sec:Binary decoder design}
\subsection{Two's complement analysis}
According to (\ref{eq:Type I_PE_LLR}), Type I PE will result in $0$ if LLRs are quantized to $\pm1$. In other words, data transferred between entities in different levels are not completely in binary form and hence can not be represented by a single bit. Therefore, $2$-bit $2$'s complement is adopted for simplicity of logical functions and demand of indicating $3$ possible LLRs $\{0,\pm1\}$.
\subsection{Input and output analysis}
\subsubsection{Type I PE}\label{subsec:Type I I/O}
According to \cite{zhang2012reduced}, universal Type I PE based on min-sum SC algorithm is a series of half or full adder-subtractors. Calculation of LLRs in (\ref{eq:Type I_PE_LLR}) is significantly simplified under $2$-bit quantization.

Unlike universal Type I PE calculation with arbitrary inputs, binary PE has a limited input set $I = \{-1,0, +1 \}$ which exhaustively lists all possible results. Suppose $X$ and $Y$ are two $2$-bit operands, $u$ is the last decoded bit which chooses the calculation pattern, and $Z$ is the output. The mathematical function of Type I PE is
\begin{eqnarray}
    Z=
    \left\{
    \begin{aligned}
    X+Y,&\quad u=0,\\
    -X+Y,&\quad u=1.
    \end{aligned}
    \right.
    \label{eq:Type I PE Hardware}
\end{eqnarray}

\begin{table}[!t]
 \begin{minipage}{.5\linewidth}
    \centering
    \caption{Results of Type I PE.}\label{tab:type i calculation}
    \tabcolsep 10pt
    \begin{tabular}{rrrVc}
    \toprule
      u &X    & Y &Z \\
    \hline
        0     & -1    & -1    & $-2\rightarrow -1$ \\
        0     & -1    & 0     & -1 \\
        0     & -1    & 1     & 0 \\
        0     & 0     & -1    & -1 \\
        0     & 0     & 0     & 0 \\
        0     & 0     & 1     & 1 \\
        0     & 1     & -1    & 0 \\
        0     & 1     & 0     & 1 \\
        0     & 1     & 1     & $2\rightarrow1$ \\
        \hline
        1     & -1    & -1    & 0 \\
        1     & -1    & 0     & 1 \\
        1     & -1    & 1     & $2\rightarrow1$ \\
        1     & 0     & -1    & -1 \\
        1     & 0     & 0     & 0 \\
        1     & 0     & 1     & 1 \\
        1     & 1     & -1    & $-2\rightarrow-1$ \\
        1     & 1     & 0     & -1 \\
        1     & 1     & 1     & 0 \\
    \bottomrule
    \end{tabular}
 \end{minipage}
 \begin{minipage}{.5\textwidth}
    \centering
    \caption{Corresponding $2$'s complement.}\label{tab:type i 2's}
    \tabcolsep 13pt
    \begin{tabular}{rrrVc}
    \toprule
      u&  X& Y &Z \\
    \hline
       0     & 11    & 11    & 11 \\
        0     & 11    & 00    & 11 \\
        0     & 11    & 01    & 00 \\
        0     & 00    & 11    & 11 \\
        0     & 00    & 00    & 00 \\
        0     & 00    & 01    & 01 \\
        0     & 01    & 11    & 00 \\
        0     & 01    & 00    & 01 \\
        0     & 01    & 01    & 01 \\
        \hline
        1     & 11    & 11    & 00 \\
        1     & 11    & 00    & 01 \\
        1     & 11    & 01    & 01 \\
        1     & 00    & 11    & 11 \\
        1     & 00    & 00    & 00 \\
        1     & 00    & 01    & 01 \\
        1     & 01    & 11    & 11 \\
        1     & 01    & 00    & 11 \\
        1     & 01    & 01    & 00 \\
    \bottomrule
    \end{tabular}
 \end{minipage}
\end{table}

Note that the results of (\ref{eq:Type I PE Hardware}) may be $\pm 2$ and will be quantized to $\pm 1$ for simplicity of calculation. Therefore, all the possible results are listed in Table~\ref{tab:type i calculation} and we can directly focus on the input and output by transforming Table~\ref{tab:type i calculation} into $2$'s complement as shown in Table~\ref{tab:type i 2's} instead of messing with those intermediate results like $\pm 1$ or $0$. In particular, we can separate the MSB and the LSB of output $Z$ and treat this PE as a combinational logic circuit with a  $5$-bit input ($u, X_M, X_L, Y_M, Y_L$) and a $2$-bit output ($Z_M, Z_L$). Therefore, Table~\ref{tab:type i 2's} is the truth table for this logic circuit which enables us to simply build corresponding logic functions.

\subsubsection{Type II PE}\label{subsec:Type II I/O}
The architecture of Type II PE is more straightforward. With binary input, (\ref{eq:Type II_PE_Min Sum}) can be pruned to
\begin{equation}\label{eq:Type II PE Hardware}
  \mathbb{L}_{N}^{(i)} = \mathbb{L}_{N/2}^{(\frac{i+1}{2})}(y_1^{\frac {N}{2}},u_{1,o}^{i-1}\oplus u_{1,e}^{i-1})\cdot \mathbb{L}_{N/2}^{(\frac{i+1}{2})}(y_{\frac {N}{2}+1}^{N},u_{1,e}^{i-1}),
\end{equation}
without obtaining the minimum of $2$ inputs since their absolute values have already been quantized to $1$.

Considering the property of multiplication, the output will be $0$ once there exists a $0$ in $2$ inputs. Therefore, hardware architecture design can be simplified by independently considering inputs $\pm 1$. Note that both $2$'s complements of $\pm 1$ have the same LSB as $1$ and the outputs can only be $\pm 1$, which means the LSB will constantly be $1$. Therefore, we can extract the MSB to analyze the input and output (I/O). I/O analysis and the corresponding $2$'s complements have been shown in Table~\ref{tab:Type ii I/O} and \ref{tab:Type ii 2's} by adopting the method mentioned in Section \ref{subsec:Type I I/O}.

\begin{table}[!t]
 \begin{minipage}{.5\linewidth}
    \centering
    \caption{Results of Type II PE.}\label{tab:Type ii I/O}
    \tabcolsep 13pt
    \begin{tabular}{rrVr}
    \toprule
    $X$& $Y$ &$Z$ \\
    \hline
     -1 & -1 &1 \\
     -1    & 1     & -1 \\
     1     & 1     & 1 \\
     1     & -1    & -1 \\

    \bottomrule
    \end{tabular}
 \end{minipage}
 \begin{minipage}{.5\linewidth}
    \centering
    \caption{$2$'s complement of the MSB.}\label{tab:Type ii 2's}
    \tabcolsep 13pt
    \begin{tabular}{ccVc}
    \toprule
        $X_M$& $Y_M$ &$Z_M$ \\
    \hline
        1     & 1     & 0 \\
        1     & 0     & 1 \\
        0     & 0     & 0 \\
        0     & 1     & 1 \\

    \bottomrule
    \end{tabular}
 \end{minipage}
\end{table}
We can conclude from Table~\ref{tab:Type ii 2's} that the calculation of the MSB of Type II PE using $2$'s complement equals to an XOR operation. Therefore Type II PE can be pruned to an XOR operation in the MSB and a fixed $1$ in the LSB.

\subsection{Design of binary PEs}
\subsubsection{Design of binary Type I PE}
Binary Type I PE can be treated as a combinational logic circuit based on the analysis in Table~\ref{tab:type i 2's}.

In this part, variable settings in Section \ref{subsec:Type I I/O} are adopted and therefore $X$ and $Y$ are two binary input operands, the last-decoded bit $u$ is a selection bit and the output is represented by $Z$. With $2$-bit quantization for $X,Y$ and $Z$, binary Type I PE consists of $5$ inputs ($u, X_M, X_L, Y_M, Y_L$) and $2$ outputs ($Z_M, Z_L$). The logical functions are listed as follows:
\begin{itemize}
\item $u=0$
    \begin{equation}
    \begin{aligned}
        Z_M &= X_L'Y_M+X_M'Y_L+X_MY_M,\\
        Z_L &= X_L'Y_L+X_M'Y_MY_L+X_LY_L'+X_MY_M;
    \end{aligned}
    \end{equation}

\item $u=1$
    \begin{equation}
    \begin{aligned}
        Z_M &= X_M'X_LY_L'+X_MY_M',\\
        Z_L &= X_L'Y_L+X_M'Y_M+X_LY_L'+X_MY_M'.
    \end{aligned}
    \end{equation}
\end{itemize}

\begin{figure}[!t]
 \begin{minipage}[t]{0.5\linewidth}
      \centering
      \includegraphics[width=0.8\linewidth]{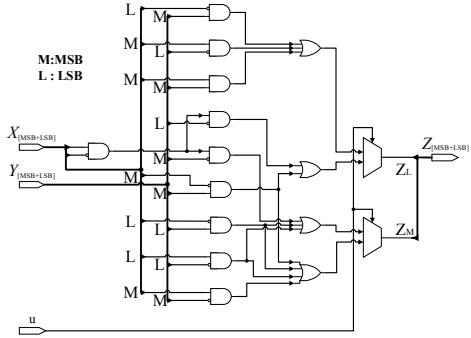}\\
      \caption{Proposed architecture of binary Type I PE.}
      \label{fig:Type I PE arch}
 \end{minipage}
 \begin{minipage}[t]{0.5\linewidth}
      \centering
      \includegraphics[width=0.99\linewidth]{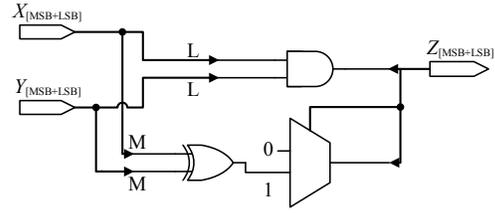}\\
      \caption{Proposed architecture of binary Type II PE.}\label{fig:Type II PE arch}
 \end{minipage}
\end{figure}

The gate-level circuit diagram of binary Type I PE is depicted in Figure~\ref{fig:Type I PE arch}.

\subsubsection{Design of binary Type II PE}
The core of Type II PE design can be concluded into $3$ key points based on the aforementioned I/O analysis.
\begin{itemize}
\itemindent 2.8em
\item[1)] MSB of Type II PE's output can be simply calculated by an XOR operation under $2$-bit $2$'s complement;
\item[2)] The LSB of Type II PE's output is fixed to $1$;
\item[3)] The output will be $0$ once there exists a $0$ in the inputs.
\end{itemize}
Architecture of binary Type II PE is shown in Figure~\ref{fig:Type II PE arch}.

\section{Performance assessment}\label{sec:simulation}
In this section, we provide the error performance of different codes and discuss their complexities.
\subsection{Settings of simulation}
We adopt a $(8192,7168)$ polar codes using different inputs under MLC NAND flash memory channels. Besides, a $(8192,7168)$ QC-LDPC code using bit-flipping decoding algorithm is also used for comparison. The selection of information length if based on \cite{mielke2008bit, takeuchi2009novel}.

We adopt a $2$-bit/cell MLC NAND flash memory model \cite{Intel97} as the simulation environment. It is assumed that the mean value of \Gaussian distribution for erase state which represents $00$ is $0$~volt and the target voltages in programming states are $3.25$ volt, $4.55$ volt, and $6.5$ volt for symbols $10$, $11$, and $01$, respectively. Standard deviations for each state are set to $2\sigma, \sigma, \sigma$, and $1.4\sigma$, where $\sigma$ changes over time due to multiple interferences. Hard-decision boundaries in binary decoder are the $3$ intersections between $4$ \Gaussian distributions, and SMMI is applied to obtain other soft-decision boundaries.

The binary-input decoder employees $2$-bit quantized LLR. Floating-point LLR is used in quantized-soft decoders. The maximum iteration is set to $15$ in hard-decision bit-flipping LDPC decoding.
\subsection{Simulation}
The result is based on FER versus raw error probability and the design of $x-$axis is explained as follows. The MLC flash memory is modeled as $4$ \Gaussian distributions and has $3$ hard-decision boundaries. In hard decoding, a raw error happens once the voltage in a \Gaussian distribution shifts to its adjacent distributions (i.e., crosses the left or right hard-decision boundary). Under \Gaussian distribution, the raw error probability $P$ can be calculated by $Q$-function.

\begin{figure}[!t]
\begin{minipage}[t]{0.5\linewidth}
  \centering
  \includegraphics[width=0.95\textwidth]{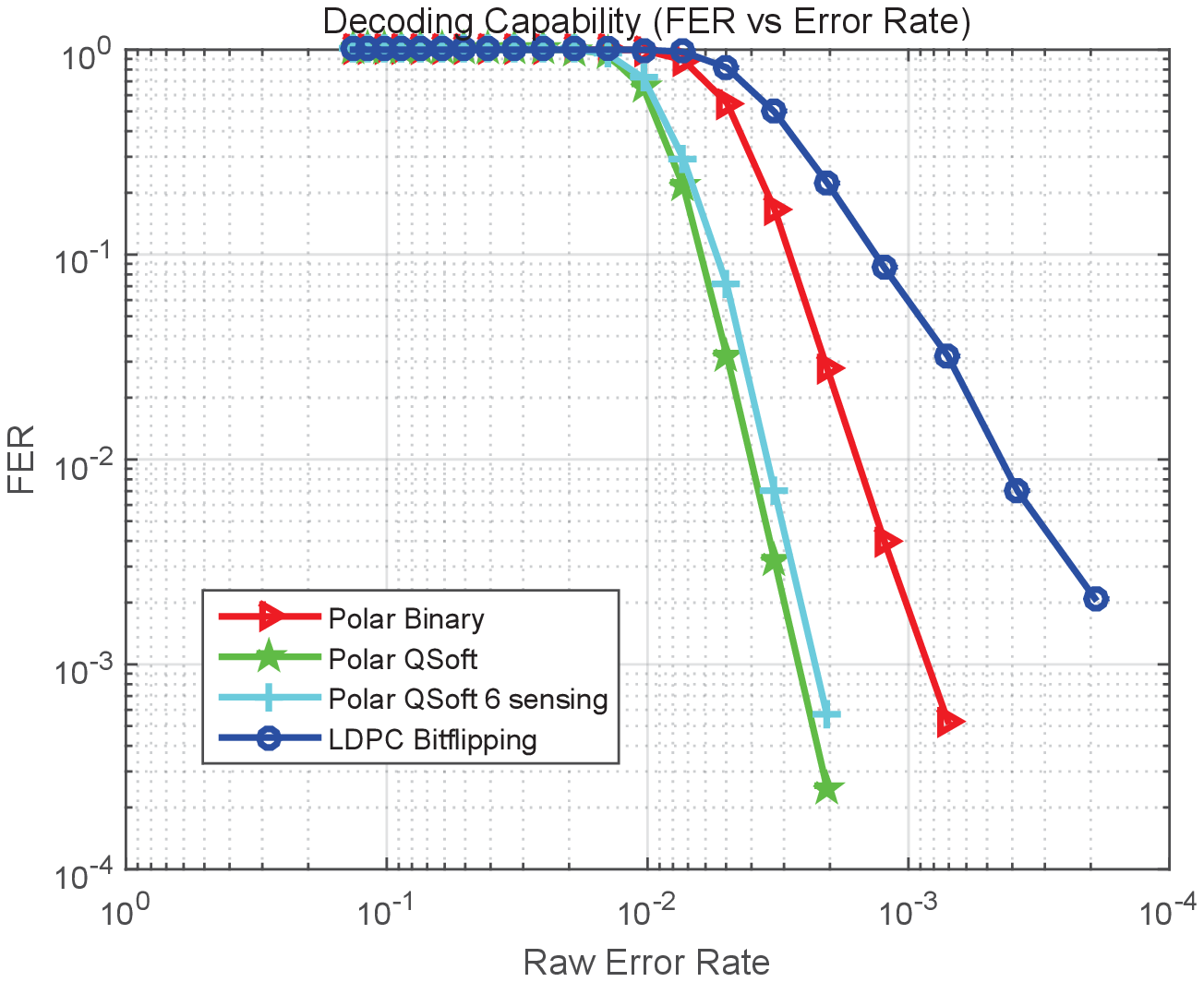}\\
  \caption{FER performance of a $(8192,7168)$ polar code and a $(8192,7168)$ QC-LDPC code.}\label{fig:FER Simulation}
\end{minipage}
\begin{minipage}[t]{0.5\linewidth}
  \centering
  \includegraphics[width=0.95\textwidth]{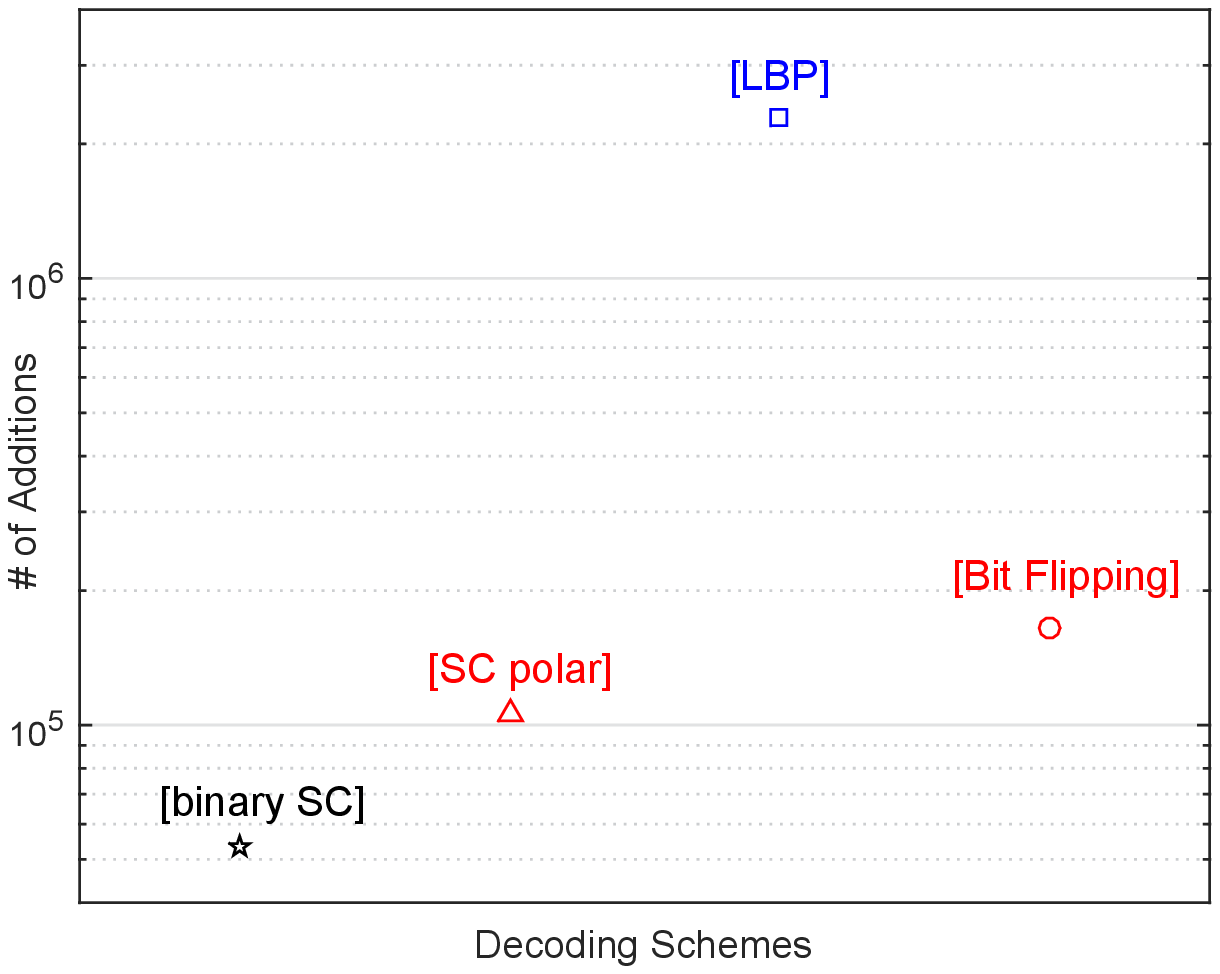}\\
  \caption{Comparison of decoding complexity between\\ different algorithms.}\label{fig:comparison}
\end{minipage}
\end{figure}

In Figure~\ref{fig:FER Simulation}, binary-input polar decoder obviously outperforms the hard-decision bit-flipping LDPC decoder. With the increment of sensing operations, quantized-soft polar codes is capable of correcting more error bits than binary-input polar code which assures the data stability of the whole system.

\subsection{Complexity analysis}
\subsubsection{Decoding of polar code}
The complexity of full size SC is $N\log_2N$, where $N$ is the code length \cite{Arikan09}. For LLR-based min-sum SC decoding, the decoder complexity is:
\begin{itemize}
  \item Type I PEs: $(N\log_2N)/2$ additions;
  \item Type II PEs: $(N\log_2N)/2$ comparisons/selection (equivalent of addition) and $(N\log_2N)/2$ sign bit multiplication (equivalent of XOR);
\end{itemize}
Overall, the decoding complexity is ($N\log_2N$) additions (XOR is negligible compared to addition).

For binary-input SC decoding, LLRs are quantized to $\pm1$, which means the comparison in Type II PEs is no longer needed. Therefore, the overall decoding complexity is $N\log_2N/2$ $2$-bit additions and $N\log_2N/2$ XOR operations.

\subsubsection{Decoding of LDPC code}
Among various LDPC decoding algorithms, min-sum algorithm is the most widely used method \cite{Xijia2015,kim2012performance,dong2011use,cui2014multilevel}. In this section, we adopt the complexity analysis of LBP decoding with min-sum algorithm in \cite{ieee802}. In this section, code length and information length are represented by $N$ and $K$. Column and row weight are denoted by $d_v$ and $d_c$.

For LBP decoding, the complexity in one iteration is:
\begin{itemize}
  \item Check node processing: $Nd_v+2(N-K)$ additions and $(2d_c-3)(N-K)+2(N-K)$ comparisons (equivalent of addition);
  \item Variable node processing: $Nd_v$ additions;
\end{itemize}
Overall, the decoding complexity is $(N-K)(2d_c+1)+2Nd_v$ additions per iteration. According to \cite{ieee802}, the LBP decoding converges within $15$ to $20$ iterations (denoted by $I$) and average column weight $\bar d_v = 3.9375$ (when code rate $R = 0.75$). To this end, $\log_2N$ is obviously smaller than $d_vI$ when $N$ is less than $8$K byte in storage system. Therefore the computational complexity of SC polar decoding is much lower than LDPC LBP decoding. The complexity of standard BP decoding with min-sum algorithm is similar to this result.

For hard-decision bit-flipping decoding, the complexity in one iteration is:
\begin{itemize}
  \item Syndrome calculation: $(N-K)d_c$ additions and multiplication in $GF(2)$;
  \item Number of unsatisfied parity checks: $Ed_c$ additions where $E$ is number of $1$'s in the syndrome;
  \item $(N-1)$ comparisons (equivalent of additions) to obtain the largest number of unsatisfied parity checks.
\end{itemize}
The complexity for bit-flipping decoding is mainly determined by the $(N-1)$ comparisons. Hence the overall decoding complexity is $I(N-1)$ additions in the worst case. In \cite{Xijia2015}, the iteration of modified gradient descent bit-flipping (MGDBF) decoder is set to $30$.
\subsubsection{Comparison of decoding complexity}
When setting code length $N$ as $8192$, information length $K$ as $7372$, iteration $I$ as $20$, column weight $d_v$ as $4$, and row weight $d_c$ as $30$, the the decoding complexity is compared in Figure~\ref{fig:comparison}. It is obvious that the proposed binary-input SC decoder has the lowest complexity. Moreover, polar codes using SC algorithm have much lower computational complexity compared to traditional LDPC codes using LBP decoding.

\section{Conclusion}\label{sec:conclusion}
This paper demonstrates that polar coded scheme holds great promise for data stability of MLC NAND flash memory. The proposed multi-strategy pre-check scheme can well balance the error performance and decoding latency. The binary-input decoder is also proposed to relieve the quantization burden of quantized-soft decoder, and lower the computational complexity compared to LDPC codes. Third, a new method named SMMI is proposed to calculate quantization boundaries without boundary searching. Finally, the \Gray code has been proved the optimal mapping scheme in our system.



\newpage
\begin{appendix}
\section{Proof for \gray code mapping scheme}\label{appd:proof for gray code}
\lemma{\Gray code can achieve best coding gain compared to any other mapping schemes.}

\proof
As mentioned in Section \ref{subsec:Gray}, we have noticed that most raw errors happen when a voltage is mistaken for its adjacent levels. Therefore, we can focus on overlapped regions when talking about mapping schemes. For the convenience of discussion, we use $4$ column vectors to indicate $4$ different states in a $2$-bit memory cell namely $A = \left( \begin{array}{l}0\\0\end{array} \right)$,
$B = \left( \begin{array}{l}1\\0\end{array} \right)$, $C = \left( \begin{array}{l}1\\1\end{array} \right)$, $D = \left( \begin{array} {l} 0\\1\end{array} \right)$. By making a full permutation for these states, we get 24 different schemes as shown in Table~\ref{tab:Gray total}.
\begin{table}[!t]
\begin{minipage}[c]{.5\linewidth}
\centering
\caption{24 Different Schemes and Number of Changes.}\label{tab:Gray total}
\tabcolsep 3pt
\begin{tabular}{*{6}{c p{4em}}}
    \bottomrule \rowcolor{lightgray! 50}
    \cellcolor {yellow}{DCBA 3} &\cellcolor{yellow} {DABC 3} &CBDA 4 & \cellcolor{yellow} {BCDA 3} &\cellcolor{yellow}{BADC 3}& \cellcolor {yellow} {ABCD 3} \\
    \hline
    DCAB 4 & DACB 4 &\cellcolor {yellow} CBAD 3 & BCAD 4 & BACD 4 & ABDC 4 \\
    \hline \rowcolor{lightgray!50}
    DBCA 5  &CDBA 4 & CABD 5 & BDCA 5 & ACBD 5 & ADBC 4\\
    \hline
    DBAC 5  &\cellcolor {yellow}CDAB 3  & CADB 5 & BDAC 5 & ACDB 5 &\cellcolor {yellow} ADCB 3\\
    \toprule
\end{tabular}
\end{minipage}
\begin{minipage}[c]{.6\linewidth}
\centering
\caption{Example of \textit{Gray} Mapping Scheme.}
\tabcolsep 5pt
\begin{tabular}{cVcccc}
\toprule
    \bfseries MSB& 0 & 1 & 1 & 0 \\
\hline
    \bfseries LSB& 0 & 0 & 1 & 1 \\
\bottomrule
\end{tabular}\label{tab:Gray schm}
\end{minipage}
\end{table}

Each scheme has $2$ rows and if $1$ bit is different from its adjacent bits in a row, we will call it a change. For example, the combination $ABCD$ indicates a mapping scheme shown in Table~\ref{tab:Gray schm}. In this case, the number of changes is $3$. The statistical results are shown in Table~\ref{tab:Gray total} . We can conclude by enumeration that the number of changes is 3 if and only if the mapping scheme is in \Gray code. The other alternatives' number of changes are $4$ or $5$.

We have already known that raw errors usually happen in overlapped regions. For a $2$-bit cell, there remains $3$ overlapped regions. Assume the raw error probability for each region is $P_1$,$P_2$ and $P_3$ respectively. Therefore, the expectation of raw errors $N_G$ of mapping schemes using \Gray code is
\begin{equation}\label{eq:Gray error P}
N_G=P_1+P_2+P_3,
\end{equation}
whereas the expectation of all other alternatives is
\begin{small}
\begin{equation}\label{eq:Other error P}
\begin{aligned}
N_A = \alpha {P_1} + \beta {P_2} + \gamma {P_3} \quad (\alpha  + \beta  + \gamma  = 4\text{ or }5,\quad\alpha \beta \gamma\neq 0).
\end{aligned}
\end{equation}
\end{small}
$N_A$ is absolutely bigger than $N_G$. In other words, we can tell that \Gray code is the best choice for mapping schemes.

\section{Calculation of boundaries in quantized-soft decoder}\label{appd:derivation}
In Section \ref{subsec:q-soft} we have mentioned that the derivation for (\ref{eq:Bond}) is wrong in \cite{kim2012performance}. This section will re-derive (\ref{eq:Bond.1}).

Since $p^{k}(x)$ is a \Gaussian distribution, $p^{(k)}(B_l^{(k)})$ and $p^{(k+1)}(B_l^{(k)})$ are
\[
\left\{ \begin{array}{l}
{p^{(k)}}(B_l^{(k)})\quad= \dfrac{1}{{{\sigma _k}\sqrt {2\pi } }}\exp ( - \dfrac{{{{(B_l^{(k)} - {\mu _k})}^2}}}{{2\sigma _k^2}}),\\
{p^{(k + 1)}}(B_l^{(k)}) =\dfrac{1}{{{\sigma _{k + 1}}\sqrt {2\pi } }}\exp ( - \dfrac{{{{(B_l^{(k)} - {\mu _{k + 1}})}^2}}}{{2\sigma _{k + 1}^2}}).
\end{array} \right.
\]
Therefore, the fraction in the left will be expanded below
\[\frac{{{\sigma _k}}}{{{\sigma _{k + 1}}}}R = \exp ( - \frac{{{{(B_l^{(k)} - {\mu _k})}^2}}}{{2\sigma _k^2}} + \frac{{{{(B_l^{(k)} - {\mu _{k + 1}})}^2}}}{{2\sigma _{k + 1}^2}}).\]
Take the log of both sides of the equation, we get
\[\log(\frac{{{\sigma _k}}}{{{\sigma _{k + 1}}}}R) =  - \frac{{{{(B_l^{(k)} - {\mu _k})}^2}}}{{2\sigma _k^2}} + \frac{{{{(B_l^{(k)} - {\mu _{k + 1}})}^2}}}{{2\sigma _{k + 1}^2}}.\]
Multiply $\sigma _k^2\sigma _{k + 1}^2$ on both sides to remove the denominator, then we have
\begin{equation*}
2\textcolor{red}{\sigma _k^2\sigma _{k + 1}^2}\log (\frac{{{\sigma _k}}}{{{\sigma _{k + 1}}}}R) = - \sigma _{k + 1}^2{({B_l}^{(k)} - {\mu _k})^2}+ \sigma _k^2{({B_l}^{(k)} - {\mu _{k + 1}})^2},
\end{equation*}
as shown in (\ref{eq:Bond}).

The other fraction can be expanded in the same way.
\end{appendix}


\begin{thebibliography}{99}
\bibitem{li2010improving} S.~Li, T.~Zhang. Improving multi-level NAND flash memory storage reliability using concatenated bch-tcm coding. IEEE Trans. {VLSI} Syst., 2010, vol.~18, no.~10, pp. 1412--1420,.

\bibitem{Kim2012low} J.~Kim, W.~Sung. Low-energy error correction of NAND flash memory thourgh soft-decision decoding. EURASIP Journal on Advances in Signal Processing, 2012, vol.~2012.

\bibitem{Grupp2012bleak}LM.~Grupp, JD. Davis, S. Swanson. The bleak future of NAND flash memory. In: Proceedings of the 10th USENIX conference on File and Storage Technologies, 2012.

\bibitem{Tutorial2013}J.~Bellorado, E.~Yaakobi. Signal Processing and Coding for Non-Volatile Memories. \url{faculty.cse.tamu.edu/ajiang/NVMW_Tutorial.eps}, 2013.

\bibitem{multi1}G. Marotta, A.~Macerola, A.~D¡¯Alessandro, et al. A 3bit/cell 32Gb NAND flash memory at 34nm with 6MB/s program throughput and with dynamic 2b/cell blocks configuration mode for a program throughput increase up to 13MB/s. In: Solid-State Circuits Conference Digest of Technical Papers (ISSCC), 2010.

\bibitem{multi2}Y.~Li, S.~Lee, Y.~Fong, et al. A 16 Gb 3-bit per cell (X3) NAND flash memory on 56 nm technology with 8 MB/s write rate. IEEE Journal of Solid-State Circuits, 2009, vol.~44(1), pp.~195-¨C207.

\bibitem{multi3}N.~Shibata, H~Maejima, K.~Isobe K, et al. A 70 nm 16 Gb 16-level-cell NAND flash memory. IEEE Journal of Solid-State Circuits, 2008, vol.~43(4), pp.~929¨C-937.

\bibitem{multi4}C.~Trinh, N.~Shibata, T.~Nakano, et al. A 5.6 MB/s 64Gb 4b/cell NAND flash memory in 43nm CMOS. In: Solid-State Circuits Conference-Digest of Technical Papers(ISSCC), 2009.

\bibitem{Xijia2015} K. C. Ho, C. L. Chen, Y. C. Liao, H. C. Chang, and C. Y. Lee, A 3.46 gb/s (9141, 8224) LDPC-based ECC scheme and on-line channel estimation for solid-state drive applications. In: Proceedings of {IEEE} Int. Symp. Circuits and Systems (ISCAS), Lisbon, Portugal, 2015.

\bibitem{dong2011use} G.~Dong, N.~Xie, and T.~Zhang, On the use of soft-decision error-correction codes in NAND flash memory. {IEEE} Trans. Circuits Syst. {I}, 2011.

\bibitem{kim2012performance} J.~Kim, D.-h. Lee, W.~Sung. Performance of rate 0.96 (68254, 65536) EG-LDPC code for NAND flash memory error correction. In: Proceedings of IEEE International Conference on Communications (ICC), Ottawa, Canada, 2012.

\bibitem{cui2014multilevel}Z.~Cui, Z.~Wang, X.~Huang. Multilevel error correction scheme for MLC flash memory. In: IEEE International Symposium on Circuits and Systems (ISCAS), 2014.

\bibitem{chen2008error}B.~Chen, X.~Zhang, Z.~Wang. Error correction for multi-level NAND flash memory using Reed-Solomon codes. In: Proceedings of IEEE Workshop on Signal Processing Systems (SiPS), 2008.

\bibitem{Arikan09} E.~Ar\i kan. Channel polarization: A method for constructing capacity-achieving codes for symmetric binary-input memoryless channels. IEEE Trans. Inf. Theory, 2009, vol.~55, no.~7, pp. 3051--3073.

\bibitem{zhang2012reduced} C.~Zhang, B.~Yuan, K.~K. Parhi. Reduced-latency sc polar decoder architectures. In: Proceedings of IEEE International Conference on Communications (ICC), 2012, pp. 3471--3475.

\bibitem{zhang2013low} C.~Zhang, K.~K. Parhi. Low-latency sequential and overlapped architectures for successive cancellation polar decoder. IEEE Trans. Signal Process., 2013, vol.~61, no.~10, pp. 2429--2441.

\bibitem{3gpp} MCC Support. Final Report of 3GPP TSG RAN WG1 $\#87$. In: 3GPP TSG WG1 Meeting $\#87$, \url{www.3gpp.org/ftp/tsg_ran/WG1_RL1/TSGR1_87/Report/}, 2016.

\bibitem{Li2015polar}Y.~Li, H.~Alhussien, E.~Haratsch, et al. A study of polar codes for MLC NAND flash memories. International Conference on Computing, NETWORKING and Communications, 2015.

\bibitem{leroux2011hardware} C.~Leroux, I.~Tal, A.~Vardy, et al. Hardware architectures for successive cancellation decoding of polar codes. In: Proceedings of IEEE International Conference on Acoustics, Speech and Signal Processing (ICASSP), Prague, Czech, 2011, pp. 1665--1668.

\bibitem{Intel97} G.~Atwood, A.~Fazio, D.~Mills, B.~Reaves. Intel strataflash memory technology overview. Intel Technology Journal, 1997.

\bibitem{cai2012error} Y.~Cai, E.~F. Haratsch, O.~Mutlu K.~Mai. Error patterns in MLC nand flash memory: Measurement, characterization, and analysis. In: Proceedings of Conference on Design, Automation and Test in Europe£¬ 2012, pp. 521--526.

\bibitem{song2016polar} H.~Song, C. Zhang, S. Zhang, et al. Polar code-based error correction code scheme for NAND flash memory applications. In: Proceedings of International Conference on Wireless Communications and Signal Processing (WCSP), 2016.

\bibitem{wang2011soft} J. Wang, T. Courtade, H. Shankar, et al. Soft information for LDPC decoding in flash: mutual-information optimized quantization. In: Proceedings of IEEE Global Telecommunications Conference (GLOBECOM), 2011.

\bibitem{wjd2012mutual} J. Wang, G. Dong, T. Zhang, et al. Mutual-information optimized quantization for LDPC decoding of accurately modeled flash data. arXiv:1202.1325, 2012.

\bibitem{mielke2008bit}N. Mielke, T. Marquart, N. Wu, et al. Bit error rate in nand flash memories. In: Proceedings of IEEE International Reliability Physics Symposium (IRPS), 2008, pp. 9¨C19.

\bibitem{takeuchi2009novel}K. Takeuchi. Novel co-design of nand flash memory and nand flash controller circuits for sub-30 nm low-power high-speed solid-state drives (ssd). IEEE Journal of Solid-State Circuits, vol. 44, no. 4, pp. 1227¨C1234, 2009.

\bibitem{ieee802} Y. Blankenship, S. Kuffner. LDPC decoding for 802.22 standard. IEEE P802.22, 2007.

\bibitem{Pan18}Q. Xu, Z. Pan, N. Liu, et al. A complexity-reduced fast successive cancellation list decoder for polar codes. Science China Information Sciences, 2018, vol. 61, no.2: pp. 022309.

\bibitem{Chen19}Z. Chen, L. Yin, Y. Pei, et al. CodeHop: Physical layer error correction and encryption with LDPC-based code hopping. Science China Information Sciences, 2016, vol. 59, no.10, pp.102309.

\end{thebibliography}
\end{document}